\documentclass[a4paper]{scrartcl}
\usepackage[T1]{fontenc}
\usepackage[utf8]{inputenc}
\usepackage{lmodern}
\usepackage[english]{babel}
\usepackage[pdftex]{graphicx}
\usepackage{amsmath,amssymb,amsthm}
\usepackage{mathrsfs}
\usepackage{mathtools}
\usepackage{grffile}
\usepackage{authblk}

\title{Unusual synchronization phenomena during electrodissolution of
silicon: the role of nonlinear global coupling}

\author[1,2]{Lennart Schmidt}
\author[1]{Konrad Sch\"{o}nleber}
\author[1]{Vladimir Garc\'{i}a-Morales}
\author[1]{Katharina Krischer}

\affil[1]{Non-Equilibrium Chemical Physics - TU M\"{u}nchen \\
  James-Franck-Str. 1, D-85748 Garching, Germany\\
  krischer@ph.tum.de \\}

\affil[2]{Institute for Advanced Study - TU M\"{u}nchen, \\
  Lichtenbergstr. 2a, D-85748 Garching, Germany}

\date{}

\begin{document}

\maketitle






\section{Introduction}
Complex spatiotemporal pattern formation is an intriguing phenomenon
often observed in nature. An example in biological systems is the spontaneous emergence of plane and
spiral calcium waves in stimulated Xenopus oocytes
\cite{Lechleiter_Science_1991}. Even human-body related phenomena are
known, such as eletrical turbulence in the heart muscle
\cite{Winfree_Science_1994, CherryFenton_NewJPhys_2008}. Spiral waves
and their break-up could be observed in cardiac tissue and these
phenomena could be explained from a solely dynamical point of
view. Thus, for example, the spiral break-up is not triggered by
inhomogeneities in the system, but rather arises via a dynamic
instability. This immediately shows that there is a need for model
systems to study the nonlinear dynamics in complex, pattern forming
systems. Such a model system is the catalytic oxidation of carbon
monoxide on a platinum surface in the UHV \cite{Ertl_Science_1991} exhibiting a
variety of patterns, like spiral waves, pulses, solitons, target
patterns and turbulence. Another model system is the
Belousov-Zhabotinsky reaction, which became a prototypical system for
the study of spiral dynamics, but gives also rise to standing,
irregular and localized clusters under global feedback \cite{Vanag_Nature_2000,
  Epstein_JPC_1996}.

In this Chapter we present spatiotemporal dynamics found
during the photoelectrodissolution of n-doped silicon and show that
this experimental system is a convenient model system to study
nonlinear dynamics with a conserved quantity: the mean-field of the
two-dimensional oscillatory medium exhibits harmonic oscillations with
constant amplitude and frequency. We model the experiments with an
adapted version of the complex Ginzburg-Landau equation
\cite{AransonKramer_RevModPhys_2002, Kuramoto_2003}, which is the
appropriate normal form close to a supercritical Hopf bifurcation. In
order to capture the mean-field oscillations, one has to introduce a
nonlinear global coupling into the equation. As we will show in what
follows, this nonlinear global coupling and the resulting conservation
law are the origin of a wide variety of spatiotemporal patterns. Even
the coexistence of regions exhibiting distinct dynamical behaviors is
observed for many parameter values.

\section{Experimental system}
\label{sec:Exp}
\begin{figure}[ht]
\centering
\includegraphics[width=\columnwidth]{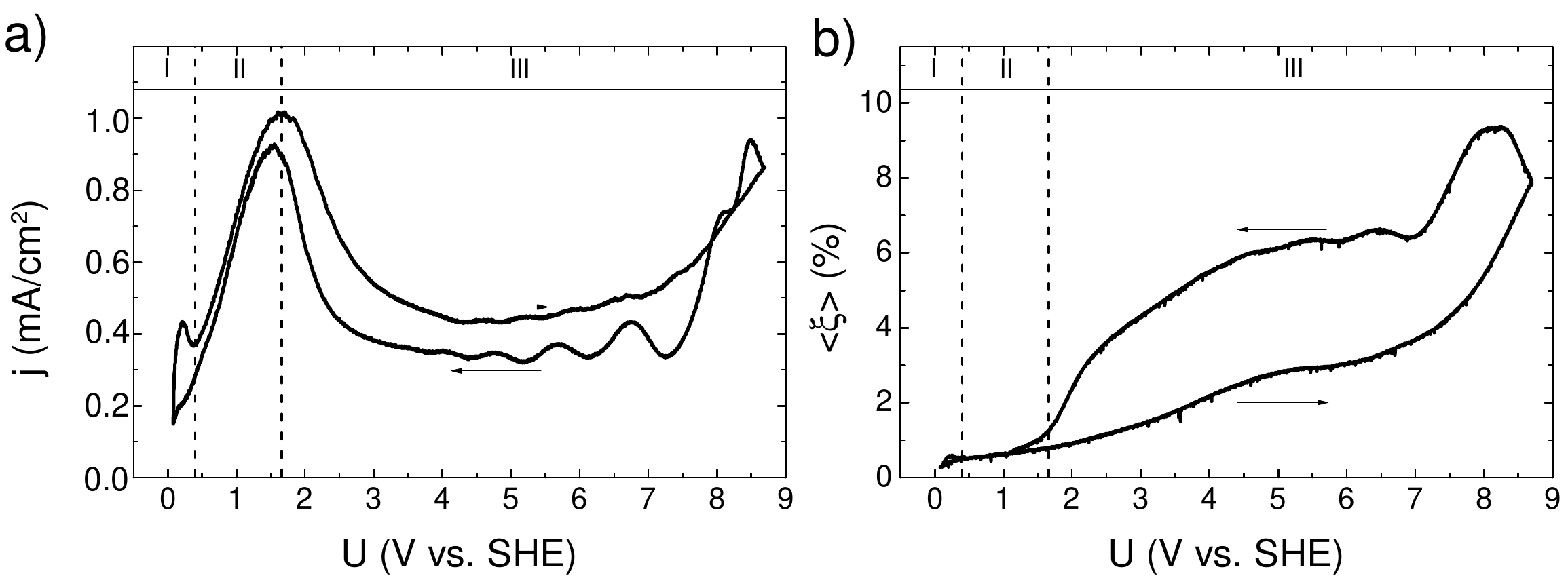}%
\caption{Cyclic voltammogram (left) (20 mV/s) of a highly illuminated n-Si sample (pH=1, $c_{\mathrm{F}}=75$ mM) and the corresponding, spatially averaged signal of the oxide layer mass $\xi$ (right). The arrows indicate the scan direction.}%
\label{fig:ExCVp}%
\end{figure}

The experimental system under consideration is the potentiostatic photoelectrodissolution of n-doped silicon under high anodic voltage in the presence of a fluoride containing electrolyte. In this process, the silicon is first electrochemically oxidized and the oxide layer is subsequently etched away purely chemically by the fluoride in the electrolyte. As a result of this interplay, depending on the external voltage, a stable oxide layer may form.\\
The anodic oxidation follows either a tetravalent or a divalent mechanism, where in both cases the first stage is an electrochemical and the second stage a chemical process \cite{Turner_JES_1958, Memming_SurfScience_1966}:
\begin{equation}
\begin{split}
\mathrm{Si+4H_2O}+\nu_{\mathrm{VB}}h^+&\rightarrow \mathrm{Si(OH)_4+4H^+}+(4-\nu_{\mathrm{VB}})e^-\\
\mathrm{Si(OH)_4}&\rightarrow\mathrm{SiO_2+2H_2O}~~~~\mathrm{(tetravalent)}\\\\
\mathrm{Si+2H_2O}+\nu_{\mathrm{VB}}h^+&\rightarrow \mathrm{Si(OH)_2+2H^+}+(2-\nu_{\mathrm{VB}})e^-\\
\mathrm{Si(OH)_2}&\rightarrow\mathrm{SiO_2+H_2}~~~~\mathrm{(divalent)}
\end{split}
\label{eq:OxMech}
\end{equation} 
Here $\nu_{\mathrm{VB}}$ is the amount of charge carriers stemming from the valence band of the silicon. As the divalent oxidation mechanism is accompanied by H$_2$ evolution the relative prevalence of both oxidation valences can be well distinguished. A significant contribution of the divalent mechanism is only found for relatively low anodic voltages \cite{Blackwood_ElectrochimActa_1992} and the reaction valency $\nu$ in the parameter regime considered in our work is close to $\nu=4$ \cite{Schoenleber_ChemPhysChem_2012}. We will thus neglect the divalent oxidation pathway for the rest of this article.\\
The initial charge transfer step for the electrochemical oxidation is
always the capture of a hole from the valence band of the silicon
leading to $\nu_{\mathrm{VB}}\geq1$ in Eq.(\ref{eq:OxMech}) \cite{Hasegawa_JES_1988}. In n-type silicon these holes have to be
photo-generated. Electron injections into the conduction band can lead
to an overall current higher than the one induced by the photon flux
incident on the surface, i.e., to external quantum efficiencies larger than one \cite{Matsumura_JEC_1983}. This current multiplication effect increases with decreasing illumination intensity and the limiting value of $\nu_{\mathrm{VB}}=1$ has been experimentally realized in literature \cite{Matsumura_JEC_1983, Blackwood_ElectrochimActa_1992}. While this trend is also present in our experiments, the values we typically find are in the range of $2\leq\nu_{\mathrm{VB}}\leq 4$.\\
The etching of the oxide is mainly due to the species HF and HF$_2^-$ in dimolecular processes \cite{Cattarin_JES_2000}. As the distribution of the fluorine to the species HF, HF$_2^-$ and F$^-$ is pH dependent and the rates for all dimolecular reaction pathways of the two etching species are different, the pH value as well as the total flourine concentration $c_{\mathrm{F}}$ determine the total etch rate. It is thus possible experimentally to vary the etch rate as well as the dominant etching pathway by the variation of the pH value of the solution and $c_{\mathrm{F}}$. Especially the voltage dependence of the etch rate is expected to show some variation with the distribution of the flourine to the respective solvated species as HF$_2^-$ and F$^-$ are charged but HF is not.\\
A typical cyclic voltammogram of a silicon electrodissolution process together with a measure of the total mass of the corresponding oxide layer $\xi$ is shown in Fig.~\ref{fig:ExCVp}a.
Below ca. $0.2$ V vs. SHE (part I) the divalent oxidation mechanism
dominates \cite{Turner_JES_1958, Memming_SurfScience_1966,
  Eddowes_JEC_1990}. Increasing the voltage the tetravalent mechanism
becomes dominant but no stable oxide layer forms as the etching
process is faster than the oxidation (part II). At voltages higher
than ca. $1.7$ V vs. SHE a stable oxide layer forms. This stable oxide
layer leads to a decrease in the total current (part III). Starting at
ca. $4$ V vs. SHE in the upward scan current oscillations can be
seen. These oscillations become even more pronounced upon reversal of
the scan direction and can also be seen in the mass of the oxide
layer, as measured with an ellipsometric setup (see below) and
depicted in Fig.~\ref{fig:ExCVp}b. The difference between the current in the upward and the downward scan can be well understood by the corresponding difference in the mass of the oxide layer on the electrode surface inhibiting the current.        

\subsection{Experimental setup}
\label{subsec:ExSet}
\begin{figure}[ht]%
\includegraphics[width=0.8\columnwidth]{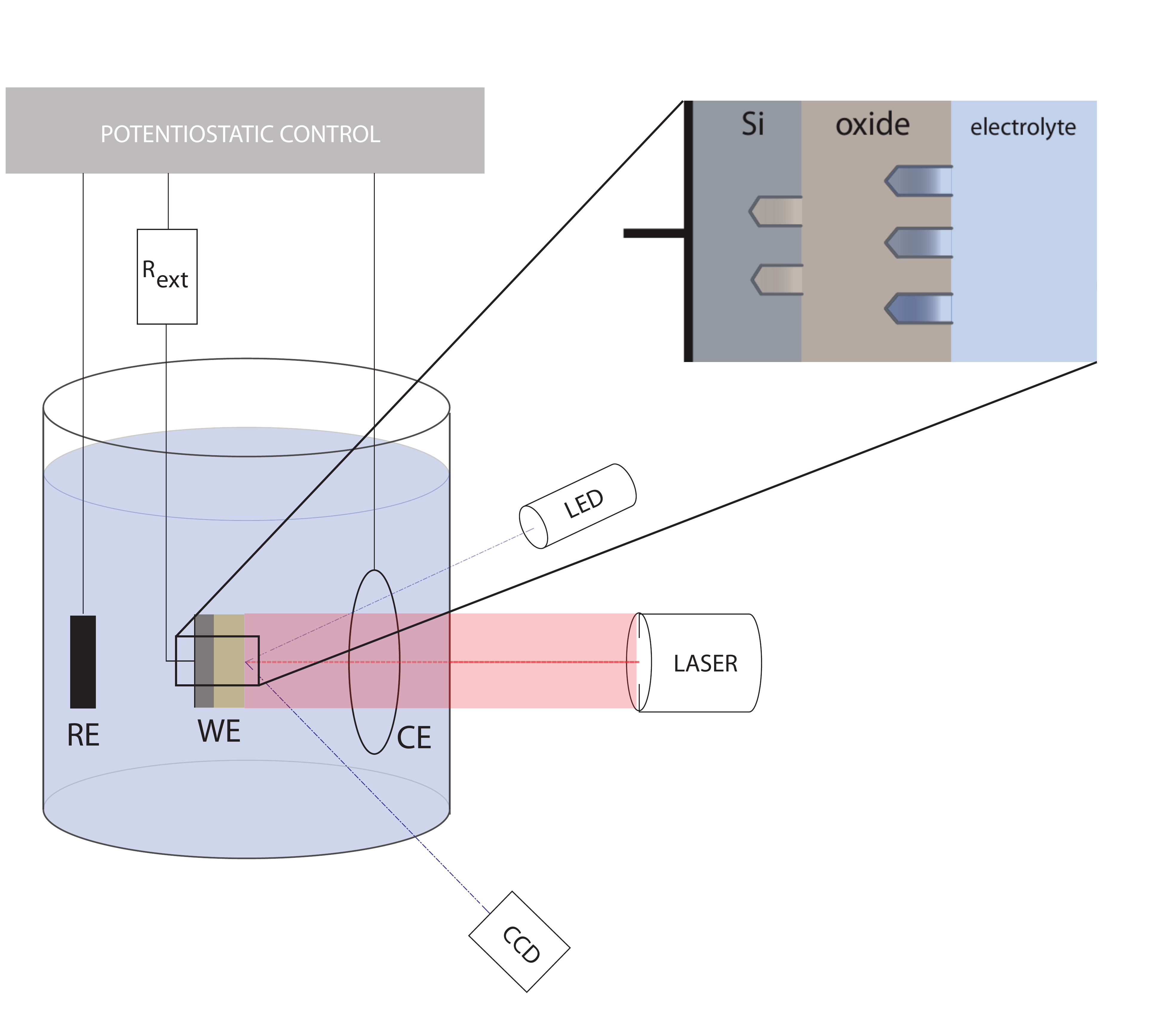}%
\caption{Sketch of the experimental setup showing the arrangement of the three electrodes and the external resistance together with the optical paths for the sample illumination (red) and the spatially resolved ellipsometric imaging (blue). A cross section of the interface and the growth direction of the oxide is shown in the inset.}%
\label{fig:ExSet}%
\end{figure}

\begin{figure}[ht!]
	\centering
		\includegraphics[width=0.8\textwidth]{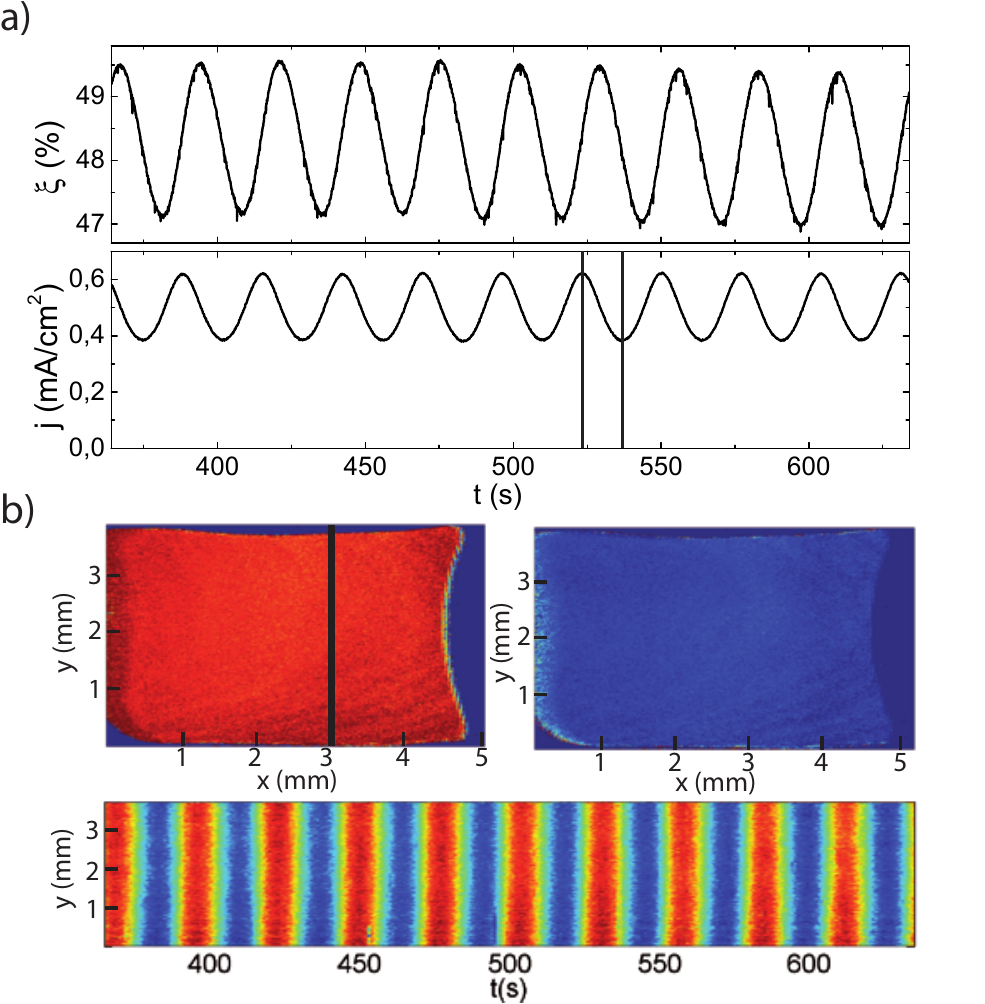}
		\caption{Uniform oscillation of $\xi$ at a high
                  illumination intensity and under constant potential
                  ($c_{\mathrm{F}}=50$ mM, $\mathrm{pH}=2.3$,
                  $R_{\mathrm{ext}}A=2.7~\mathrm{k\Omega cm^2}$,
                  $I_{\mathrm{ill}}=3.0$ mW/cm$^2$, $U=8.65$ V vs. SHE). \textbf{a)} Time
                  series of the global quantities $\xi$ and $j$;
                  \textbf{b)} Ellipsometric intensity distribution on
                  the electrode for the two times indicated in a) and
                  the temporal evolution of a 1d cut along the line
                  indicated in the left electrode picture. Red
                  indicates a relatively high and blue a relatively
                  low value of $\xi$.}
	\label{fig:NurHom}
\end{figure}

To study the properties of the oxide layer during the potentiostatic
photoelectrodissolution of n-doped silicon, we use a three electrode
electrochemical cell equipped with an ellipsometric imaging system
providing spatially resolved information on the oxide layer mass as shown in Fig.~\ref{fig:ExSet}. 
The polarization of elliptically polarized light incident on the
surface is changed upon reflection at the Si$|$SiO$_2|$Electrolyte
interface and this
change is then converted into an intensity signal by a polarizer (blue
path in Fig.~\ref{fig:ExSet}). Insertion of an imaging optic in the
reflected light path
gives an image of the electrode on a CCD chip. The system allows
the in situ measurement of the optical path through the silicon oxide
at a given area on the surface with a spatial resolution in the
$10~\mu$m range. We call the intensity at each pixel relative to the
detection limit of the CCD the ellipsometric intensity $\xi$. For an optimal contrast an angle of incidence on the sample surface close to the Brewster angle of the Si$|$Water system (ca. $70^{\circ}$) has to be chosen.\\
Silicon samples and electrolyte are prepared as described in \cite{Schmidt_Chaos_2014} and the electrolyte is kept under an argon atmosphere throughout the experiments. To minimize parameter variations across the silicon surface the electrolyte solution is constantly stirred. Furthermore, the counter electrode is placed symmetrically opposite the silicon working electrode at a distance of several centimeters to minimize possible coupling effects \cite{Mazouz_PRE_1997}. Under these conditions perfectly uniform oscillations in $\xi$ can be realized as shown in Fig.(\ref{fig:NurHom}).

\clearpage

\subsection{Dynamics}
\label{subsec:Dyn}
\begin{figure}[ht]
\centering
\includegraphics[width=11cm]{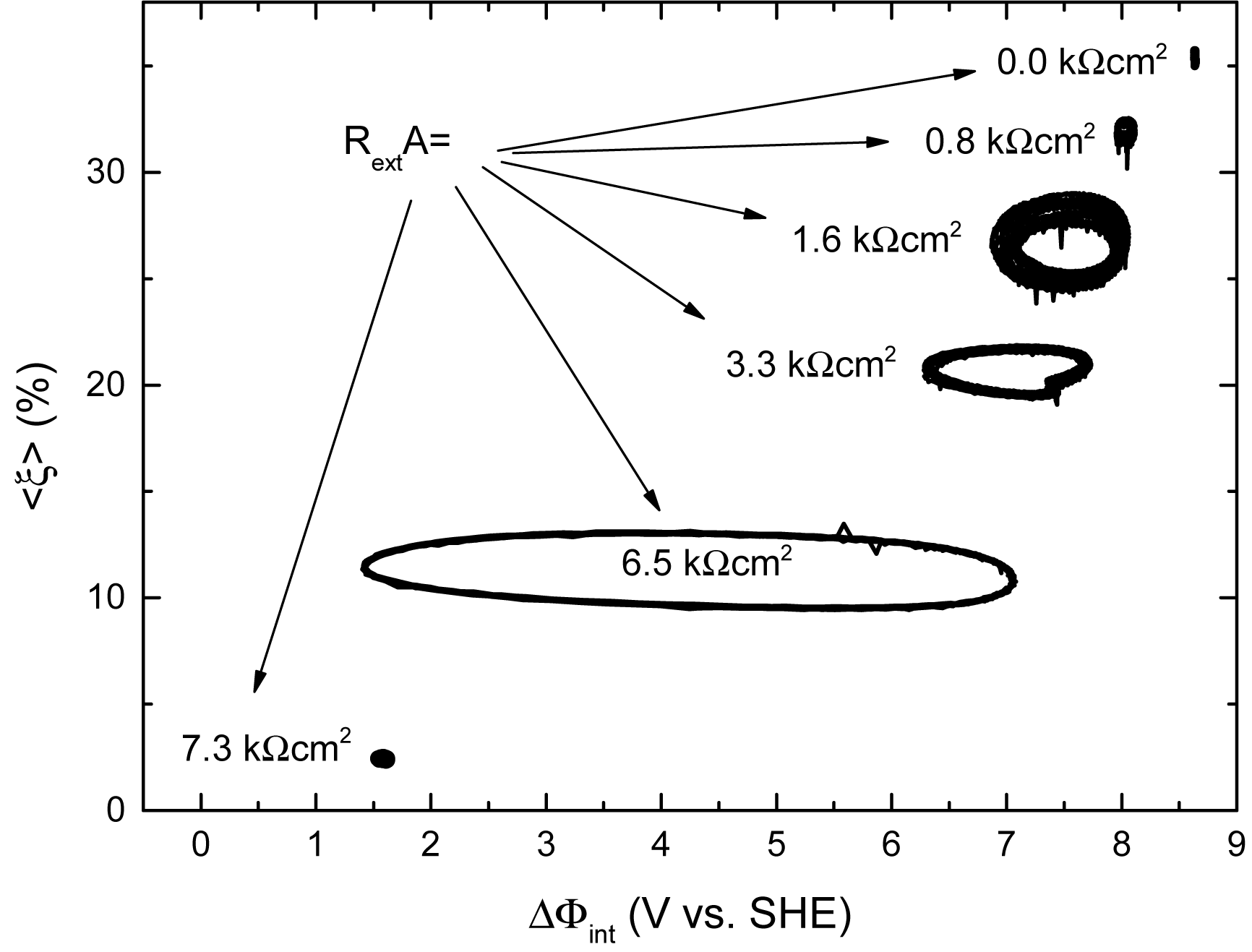}%
\caption{Phase space plots of spatially averaged time series measured
  during the photoelectrodissolution of highly illuminated n-doped
  silicon (pH 1, $c_{\mathrm{F}}=75$ mM, $U=8.65$ V vs. SHE) with varying $R_{\mathrm{ext}}$.}%
\label{fig:ExpHopf}%
\end{figure}

\begin{figure}[ht]
\centering
\includegraphics[width=\columnwidth]{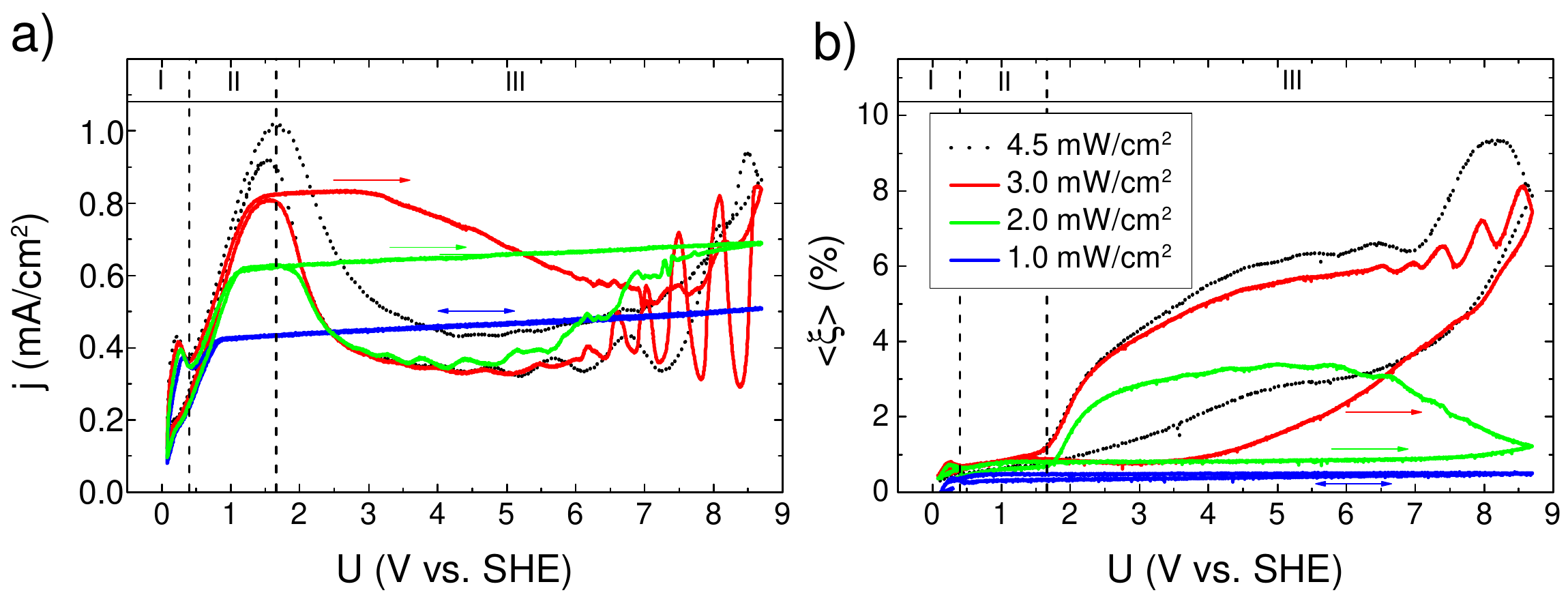}
\caption{Cyclic voltammogram (left) (20 mV/s) of an n-Si sample (pH=1, $c_{\mathrm{F}}=75$ mM) at the levels of illumination indicated and the corresponding, spatially averaged signal of the oxide layer mass $\left\langle\xi\right\rangle$ (right). The highly illuminated case (dot) is identical to Fig~\ref{fig:ExCVp}.}%
\label{fig:ExCVn}%
\end{figure}

\begin{figure}[ht]
\centering
\includegraphics[width=11cm]{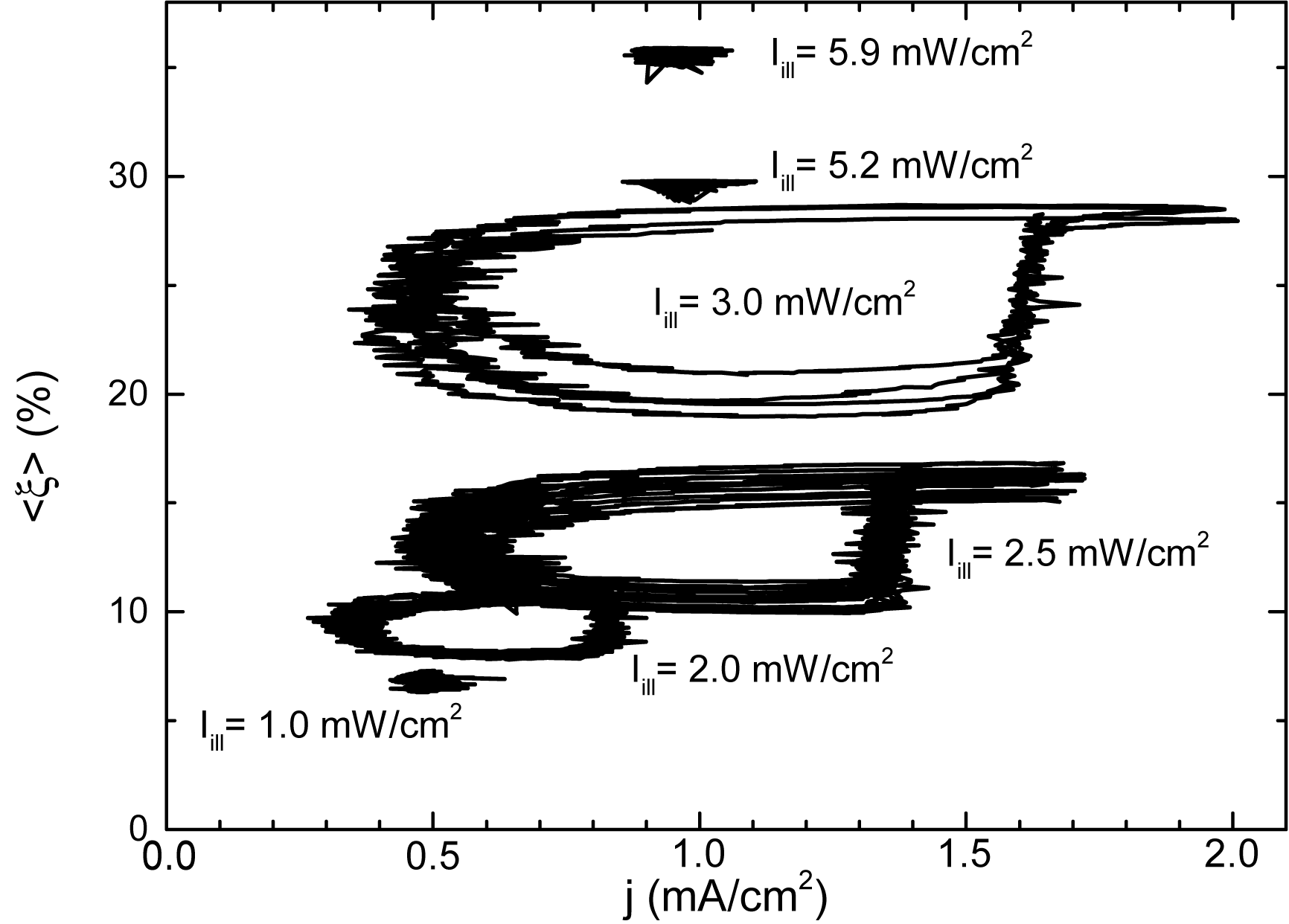}%
\caption{Phase space plots of spatially averaged time series occuring during the photoelectrodissolution of n-doped silicon at various levels of the illumination intensity $I_{\mathrm{ill}}$ (pH 1, $c_{\mathrm{F}}=75$ mM) without an external resistance.}%
\label{fig:ExHopfn}%
\end{figure}

As early as 1958 it was established that the potentiostatic
electrodissolution of p-doped silicon can proceed in an oscillatory
fashion when the applied anodic bias is sufficient
\cite{Turner_JES_1958}. To stabilize the otherwise damped oscillations
an external resistor connected in series with the working electrode
has been found to be indispensible \cite{Chazalviel:1992a}. This external
resistance $R_{\mathrm{ext}}$ links the potential drop across the
silicon$|$oxide$|$electrolyte interface $\Delta\phi_{\mathrm{int}}$ to
the total current $I$ passing through the surface.
\begin{equation}
\Delta\phi_{\mathrm{int}}=U-R_{\mathrm{ext}}\cdot I
\label{eq:GlobCoup}
\end{equation}
Thus, the external resistance introduces a coupling between all points
at the electrode surface. This coupling is both global, as only the
spatial average of the current is relevant, and linear. The behavior
of the electrodissolution of n-doped silicon for sufficiently high illumination is identical to that of p-doped silicon \cite{Paolucci:1992, Schoenleber_NJP_2014}. This can be explained by the fact that the amount of holes in the valence band of the silicon is in this case always sufficient to maintain the current determined by the electrochemical parameters. 
In both cases the oscillations arise from a Hopf bifurcation occuring at a minimal, electrolyte specific threshold value of the external resistance \cite{Miethe_JEC_2012} as shown in Fig~\ref{fig:ExpHopf}.   
It is clearly visible that with an increase of the external resistance
first stable, sinusoidal oscillations with amplitudes increasing with
the external resistance are found. This is the expected behavior close
to a Hopf bifurcation. Upon further increase of $R_{\mathrm{ext}}$ the
shape of the oscillations then gradually changes towards a more
relaxational type. Above another electrolyte specific threshold value
of $R_{\mathrm{ext}}$ the oscillations vanish abruptly and instead the
system relaxes to a stable node. The upper boundary of the oscillatory
regime with respect to the external resistance can be well understood
by comparing the voltage drop across the interface in this case as
shown in Fig.~\ref{fig:ExpHopf} to the CV scan shown in
Fig.~\ref{fig:ExCVp}. The latter shows that the value of
$\Delta\phi_{\mathrm{int}}$ for $R_{\mathrm{ext}}$ above the oscillation boundary is in the voltage region where no stable oxide can form. The extent of the oscillatory regime is thus determined by a Hopf bifurcation at the low coupling limit and a cut-off caused by leaving the experimental parameters for a stable oxide in the high coupling limit.\\\\
A second coupling mechanism can be introduced by decreasing the
illumination intensity, thus cutting off the total current caused by
limiting the generation of holes in the valence band of the n-doped
silicon samples. A lower illumination intensity leads to a higher
coupling strength. This coupling has a nonlinear characteristic and is
at least partly global as the total current, i.e. the current averaged
over all points on the surface, is again determining its strength. The
cyclic voltammogram changes significantly when this coupling is
introduced as shown in Fig.~\ref{fig:ExCVn} \cite{Eddowes_JEC_1990},
together with corresponding oxide mass changes.
Comparing the illumination limited cyclic scans to the unlimited case,
one notes that when the current reaches the illumination limit the
oxide growth is initially suppressed. Only at significantly higher
potentials does an oxide layer form. The potential shift for the oxide
formation is illumination dependent and at too low illumination levels
no oxide formation is found at all. In the cases where a stable oxide layer forms, again, both current and oxide layer mass show oscillations. The illumination limitation induced coupling is by itself sufficient to generate sustained oscillations as shown in the phase space plots in Fig~\ref{fig:ExHopfn} 
Again the coupling strength is also a bifurcation parameter leading to
stable foci below and stable oscillations above an electrolyte
specific threshold value. At very high coupling strength the system
relaxes to a stable node. In contrast to the case of the linear global
coupling discussed above, however, the transition to the steady state
is not abrupt and thus not of the same origin. It is linked to pattern
formation and can thus only be
understood considering the spatially extended system. 

\subsection{Spatially extended system}
\label{subsec:SpatExt} 
\begin{figure}[ht]
	\centering
		\includegraphics[width=0.8\textwidth]{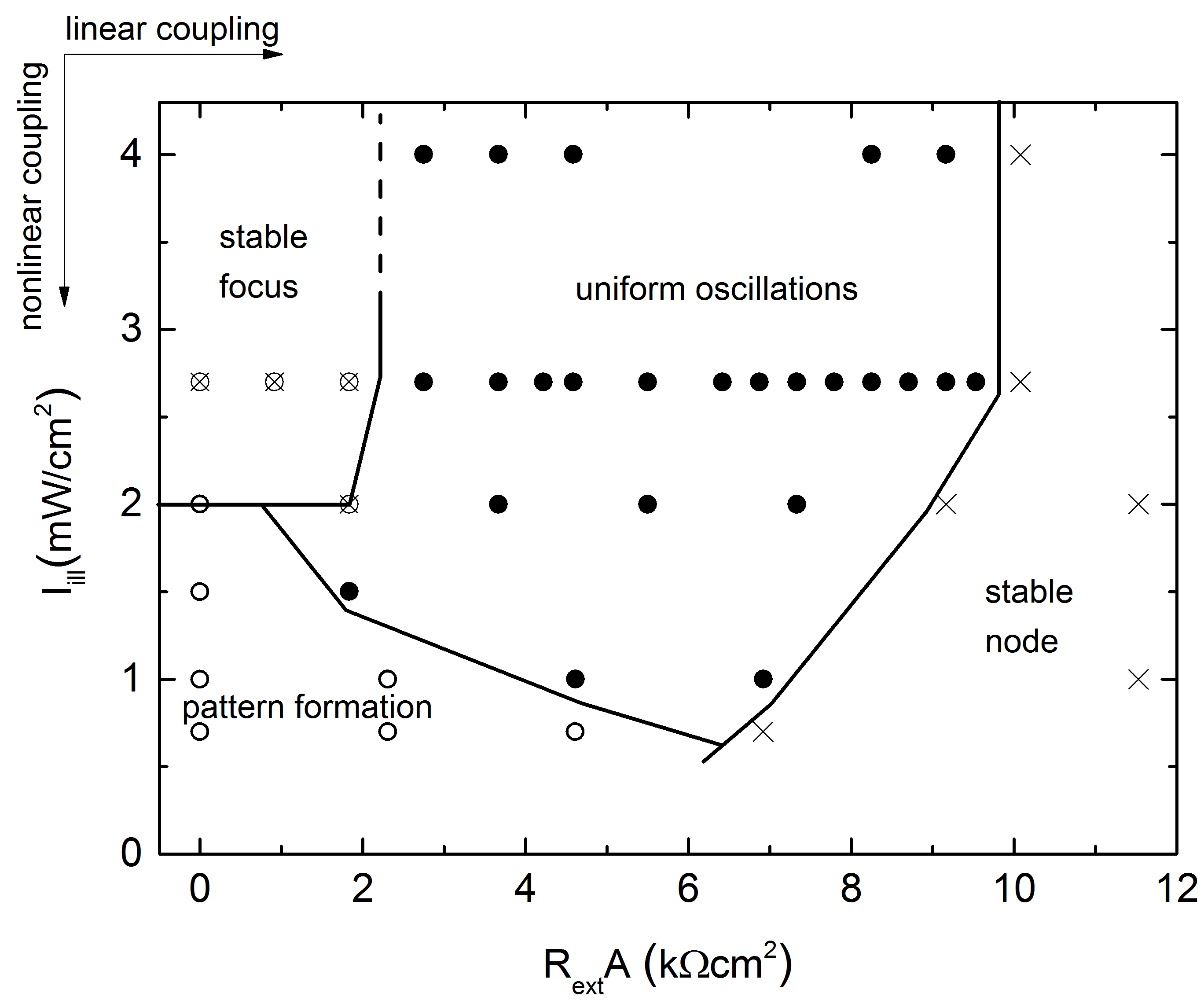}
		\caption{Dependence of the oscillation type on the strengths of the linear, global coupling $R_{\mathrm{ext}}\cdot A$ and the nonlinear coupling, the restriction of  $I_{\mathrm{ill}}$.}
	\label{fig:Para}
\end{figure}

When regarding the spatially extended system an important difference
between oscillations stabilized by the linear global coupling and
oscillations stabilized by the nonlinear coupling becomes evident. In
the former case the oscillations are always spatially uniform, while
in the latter case they are often accompanied by pattern
formation. This behavior was found in our group purely by chance as an
n-doped silicon sample was erroneously used instead of a p-doped
one. Unsurprisingly the behavior was quite unexpected and a current
could only be seen when the lightproof box in which the experiment
resided was opened. Under these conditions the patterns in $\xi$ where
first observed. 
Patterns are also found when the spatially resolved ellipsometric
intensity recorded during the cyclic voltammograms shown in
Fig.~\ref{fig:ExCVp} and Fig.~\ref{fig:ExCVn} is regarded. While in
the highly illuminated case, i.e. at negligible nonlinear coupling,
oxide growth and also the oscillations in
$\left\langle\xi\right\rangle$ are spatially uniform, the oxide growth
under restricted illumination proceeds along a growing wave front and
the oscillations show spatial patterns in $\xi$. The nonlinear
coupling is thus experimentally indispensible for pattern formation to
occur. In experiments with a constant voltage, growing wave fronts are also present in the initial transients
preceding the oscillations accompanied by pattern formation.\\
If both coupling mechanisms are combined, pattern formation is found
as long as the linear global coupling is not too strong compared to
the nonlinear coupling, leading to a typical parameter space as shown
in Fig.~\ref{fig:Para} \cite{Schoenleber_NJP_2014}   

In Section~\ref{sec:Results_and_Discussion} we will present the patterns found in the
experiments together with simulation results for a theoretical model
system described in what follows.

\section{Theoretical modelling of experiments}
The experimental system presented in the preceding section can be
modelled in a very general way. At high illumination intensities we
observe homogeneous oscillations over the entire electrode
surface. These oscillations originate in a Hopf bifurcation
as described above (see Fig.~\ref{fig:ExHopfn}). Thus, in order to model this system, the appropriate normal form to start with is the
complex Ginzburg-Landau equation (CGLE) \cite{AransonKramer_RevModPhys_2002,
  Kuramoto_2003, GarciaMorales_ContempPhys_2012} for a complex order
parameter $W(\mathbf x,t)$

\begin{equation}
  \partial_t W = W + (1 + i c_1) \nabla^2 W - (1 + i c_2)
  \left| W \right|^2 W \ .
\label{eq:CGLE}
\end{equation}

This equation describes all reaction-diffusion equations in the
vicinity of a Hopf bifurcation. For a general derivation see \cite{Kuramoto_2003}.
Equation~\eqref{eq:CGLE} admits plane wave solutions of wavenumber $Q$

\begin{equation}
  W_Q = a_Q \exp \left[ i (\omega_Q t + Qx) \right] \ ,
\end{equation}

with $\left| a_Q \right|^2 = 1-Q^2$ and $\omega_Q = -c_2 + (c_2 - c_1)
Q^2$ \cite{Kuramoto_2003, GarciaMorales_ContempPhys_2012}. A general
solution is then given as a combination of these plane waves. This, in
general, results in dynamics with an unpreserved homogeneous mode
$W_0 = \left< W \right>$. In contrast, for a huge parameter space the silicon system
exhibits conserved harmonic oscillations in the averaged oxide-layer
thickness. To achieve this in our model we adjust the CGLE in the most straightforward way by introducing a nonlinear
global coupling into Eq.~\eqref{eq:CGLE}, leading to a modified
complex Ginzburg-Landau equation (MCGLE) \cite{Miethe_PRL_2009, GarciaMorales_PRE_2010}

\begin{equation}
  \partial_t W = W + (1 + i c_1) \nabla^2 W - (1 + i c_2)
  \left| W \right|^2 W - (1 + i \nu) \left< W \right> + (1 + i c_2) \left<
    \left| W \right|^2 W \right> \ .
\label{eq:MCGLE}
\end{equation}

Since we model a two-dimensional system, the complex order parameter
$W(\mathbf r, t)$ is a function of the position vector $\mathbf r =
(x,y)$ and time $t$. Angular brackets $\left< \dots \right>$ denote
the spatial average. Now, when taking the spatial average of the whole
equation, Eq.~\eqref{eq:MCGLE}, one obtains

\begin{equation}
  \partial_t \left< W \right> = - i\nu \left< W \right> \ ,
\end{equation}

which results in conserved harmonic oscillations of the spatial
average,

\begin{equation}
  \left< W \right> = W_0 = \eta \exp\left[ -i(\nu t + \phi_0) \right]
  \ ,
\label{eq:conservation_law}
\end{equation}

with amplitude $\eta$ and frequency $\nu$. $\phi_0$ is an arbitrary
initial phase. The essential dynamical properties of the silicon
system are thus met with Eq.~\eqref{eq:MCGLE}: oscillations arising
through a Hopf bifurcation and the conserved harmonic mean-field
oscillation. In Section~\ref{sec:Results_and_Discussion} we show that
this general ansatz indeed captures the pattern dynamics found in the experiments.

\subsection{Clusters}
\label{sec:Clusters}
The common notion of phase clusters describes a state, where the oscillatory
medium separates into several parts. The oscillations in the
different parts are phase shifted with respect to each other
\cite{Vanag_Nature_2000, Vanag_JPCA_2000,
  Mikhailov_PhysicsReports_2006, Lin_PRE_2004, Kaira_PRE_2008}. In the most simple case the clusters
are arranged symmetrically and therefore the phase shifts in case of
$n$ clusters amount to $2\pi m/ n$ \cite{Okuda_PhysicaD_1993,
  Lin_PRE_2004, Kaira_PRE_2008}, where $m =
1, 2, \dots, n-1$. Typically,
in the case of cluster patterns the dynamics can be reduced to a phase
model. However, there exist a second type of clusters, where essential
variations in the amplitudes are present, called type II clusters
\cite{Varela_PCCP_2005, Miethe_PRL_2009, GarciaMorales_PRE_2010}. We will see that this second type of cluster
patterns naturally arises in our experiments and can be reproduced
with the MCGLE.

But first of all we have to clarify, how clustering can occur and why it is possible in the MCGLE. Note that in the CGLE, Eq.~\eqref{eq:CGLE}, cluster patterns
cannot emerge, since the dynamics are invariant under a phase shift $W
\rightarrow e^{i\chi} W$, for arbitrary $\chi$. 
For clustering to occur this symmetry has to be broken.
This becomes clear when considering
two-phase clusters with a period of $T_0$ and phase balance (i.e. both
clusters have the same size). Then, the dynamical equations have to
stay invariant only when shifting the time $t_0$ to $t_0 + T_0/2$, but
they are no longer invariant with respect to arbitrary shifts in
time. In terms of the complex order parameter $W$ this means that the
dynamical equations are only invariant under the discrete transformation

\begin{equation}
  W = \hat{W} \exp\left[i \omega_0 t \right] \quad \rightarrow \quad \hat{W}
  \exp \left[i \left( \omega_0 (t + T_0/2) \right) \right] \ .
\end{equation}

With $\omega_0 = 2\pi/T_0$ the transformation reads

\begin{equation}
  W \quad \rightarrow \quad e^{i\pi} W \ .
\end{equation}

In general, for $n$ clusters, one needs that the equations are
invariant under the discrete symmetry

\begin{equation}
  W \quad \rightarrow \quad e^{i 2 \pi /n} W \ .
\end{equation}

To account for this symmetry the proper extension of the CGLE is given
by the term $\gamma_n W^{*n-1}$, describing also an external resonant
forcing \cite{Gambaudo_JDE_1985, CoulletEmilsson_PhysicaD_1992, Petrov_Nature_1997,
  GarciaMorales_ContempPhys_2012, Lin_PRE_2004, Yochelis_EPL_2005, Kaira_PRE_2008,
  Conway_PRE_2007, Marts_Chaos_2006, Yochelis_PhysicaD_2004}.
The asterisk denotes complex conjugation.

We will see that such a symmetry breaking term is intrinsically
present in the MCGLE. Therefore, we write $W = W_0(1+w)$ for the
complex amplitude in Eq.~\eqref{eq:MCGLE}. By exploiting the
conservation law, Eq.~\eqref{eq:conservation_law}, and the resulting fact that
$\left< w \right> = \left< w^* \right> = 0$, one obtains again a CGLE, now for
the inhomogeneity $w$, which reads \cite{GarciaMorales_PRE_2010}

\begin{align}
  \partial_t w = &(\mu + i \beta) w + (1 + i c_1) \nabla^2 w
  \notag \\
  &-(1 + i c_2) \eta^2 (\left| w \right|^2 w + w^*) + C \ , 
\label{eq:CGLE_w}
\end{align}

where

\begin{equation}
  C = (1 + i c_2) \eta^2 \left[ \left< 2\left| w \right|^2 + w^2
    \right> - (2 \left| w \right|^2 + w^2 ) \right]
\end{equation}

and $\mu = 1 - 2\eta^2$, $\beta = \nu - 2 c_2 \eta^2$. Here the needed
symmetry-breaking term $-(1+ic_2) \eta^2 w^*$ occurs. But note that
this term does not arise from the nonlinear global coupling. It would
be present also when considering the CGLE, Eq.~\eqref{eq:CGLE},
without additional couplings. Crucial are the terms in $C$ proportional to $\left| w
\right|^2$ and $w^2$. As long as they are present, the equation is not
symmetric with respect to the transformation $w \rightarrow
e^{i\psi} w$ for any $\psi$. Here the nonlinear global coupling comes into play, as it renders a
vanishing $C$ possible, via the term proportional to $\left< 2\left| w
  \right|^2 + w^2 \right>$. For this case, i.e. for $C=0$, the
occurence of clusters is possible and the equation is symmetric with
respect to the discrete symmetry $w \rightarrow e^{i\pi} w$. Note that
this is impossible for a solely linear global coupling.

\section{Results \& Discussion}
\label{sec:Results_and_Discussion}
In the following sections we will demonstrate how well
the dynamics of the oxide-layer thickness are captured with our very
general ansatz in Eq.~\eqref{eq:MCGLE}.

\subsection{Cluster patterns}
\begin{figure}[ht]
  \centering
  \includegraphics[width=11cm]{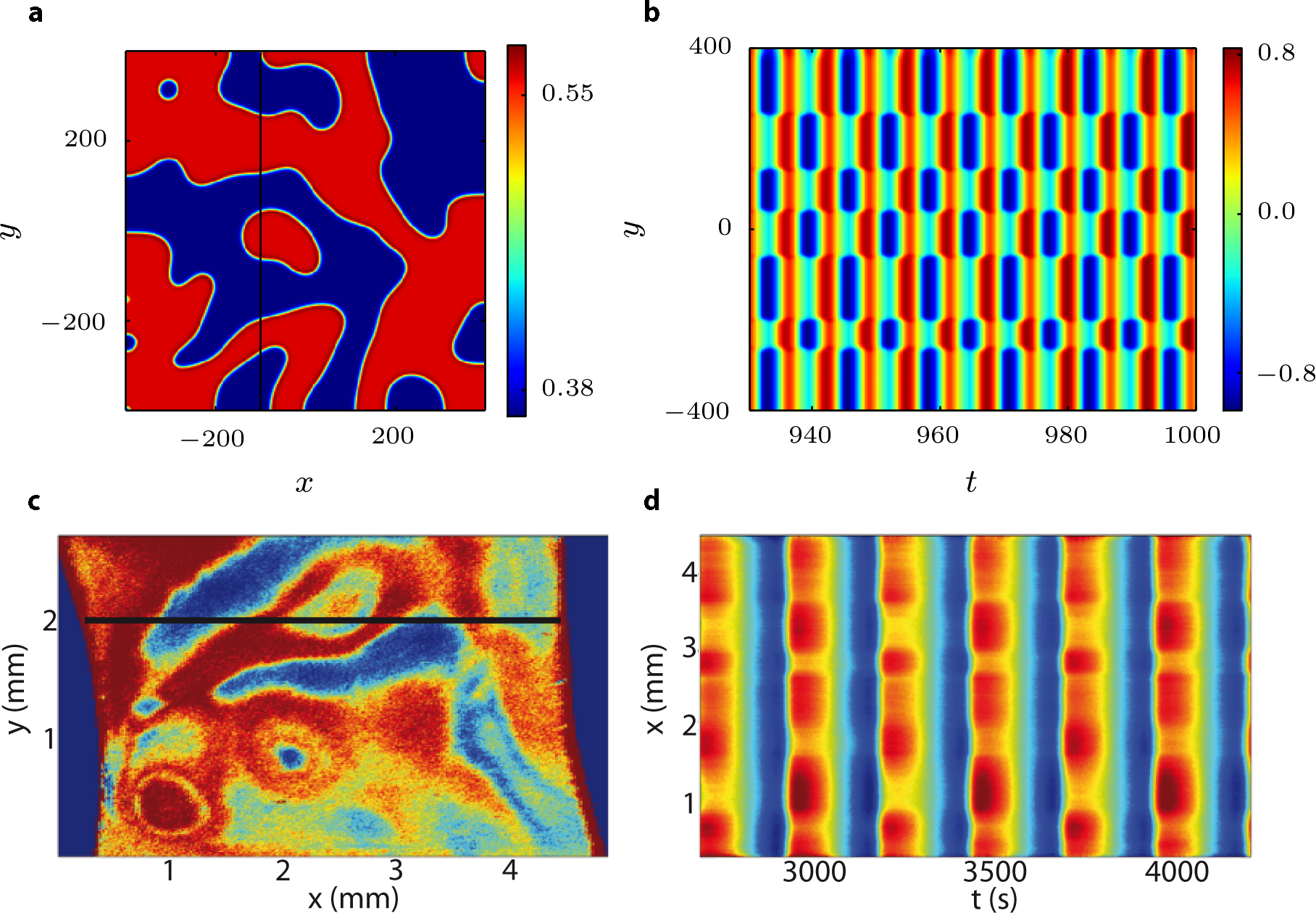}
 \caption{Two-phase clusters in theory (a,b) and experiment (c,d). (a) Snapshot
    of the two-dimensional oscillatory medium in the theory. Shown is
    the real part of $W$. (b) Spatio-temporal dynamics in an
    one-dimensional cut versus time in the theory. (c,d) The same as
    (a,b) now for the experimental results. The simulation captures
    the experimental dynamics very well. Note that the colorbars are
    different for each subfigure. Parameters read: $c_1 = 0.2$, $c_2 = -0.58$, $\nu = 1.0$, $\eta = 0.66$ (simulation) and $c_{\mathrm{F}}=35$ mM, pH=1, $R_{\mathrm{ext}}\cdot A=9.1$ k$\Omega$cm$^2$, $I_{\mathrm{ill}}=0.7$ mW/cm$^2$ (experiment).}
\label{fig:cluster}
\end{figure}

As we have clarified the theoretical basis for the emergence of
cluster patterns in Section~\ref{sec:Clusters}, we now turn towards the results of our experiments
and simulations.
In Fig.~\ref{fig:cluster} we compare the cluster dynamics in the
simulations of Eq.~\eqref{eq:MCGLE} with the experimental ones \cite{Schmidt_Chaos_2014}. In
Fig.~\ref{fig:cluster}a and c two-dimensional snapshots of the
simulations and the experiments, respectively, are shown. The
spatio-temporal dynamics can be seen in one-dimensional cuts in
Fig.~\ref{fig:cluster}b and d for the simulations and the experiments,
respectively. They show that the homogeneous oscillation is modulated
by two-phase clusters. Therefore, it is clear, that the phase shift
between two regions is not given by $\pi$.

\begin{figure}[ht]
  \centering
  \includegraphics[width=8.5cm]{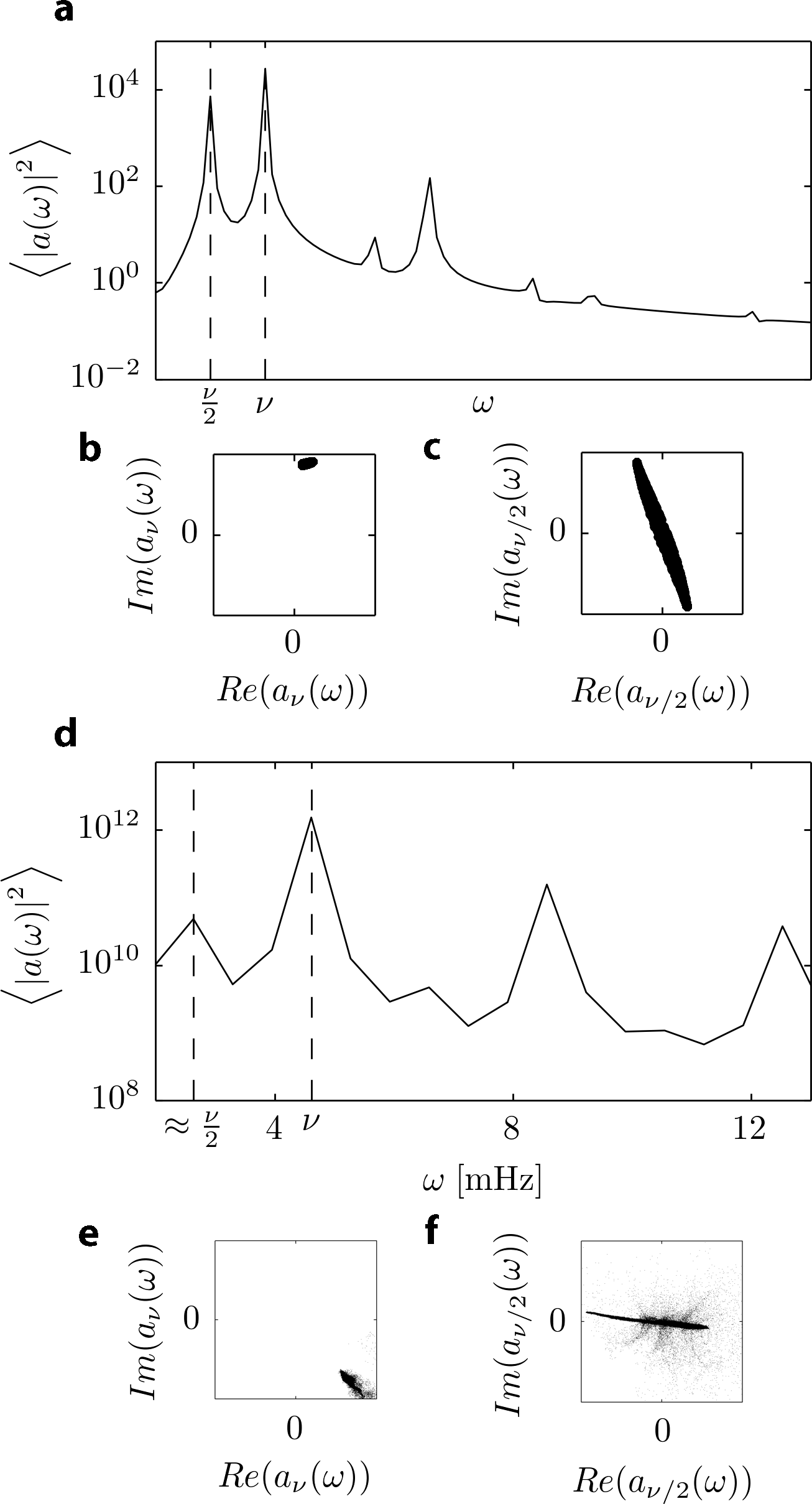}
  \caption{(a,d) Cumulative power spectra for experiments and
    simulations, respectively. The two major peaks at $\nu$ and
    approximately $\nu/2$ are indicated. The arrangement of the local
    Fourier amplitudes in the complex plane corresponding to these peaks are
    depicted in (b) and (c) for the theory and (e) and (f) for the
    experiments, respectively. The whole two-dimensional system is
    considered, which leads to the scattered oscillators in the
    experimental result in (f).}
\label{fig:Fourier_analysis_cluster}
\end{figure}

To analyze time series of this we perform a Fourier transformation in
time at every point $\mathbf r$ of the
ellipsometric signal and of the real part of $W(\mathbf r,t)$ for the
experiments and simulations, respectively \cite{Miethe_PRL_2009}. We spatially average the resulting
amplitudes $\left| a(\mathbf r, \omega) \right|^2$ to obtain the cumulative power spectrum
$S(\omega) = \left< \left| a(\mathbf r, \omega) \right|^2
\right>$. Results are shown in Fig.~\ref{fig:Fourier_analysis_cluster}.

Two major peaks occur in both cumulative power spectra in
Fig.~\ref{fig:Fourier_analysis_cluster}a (theory) and d (experiment), one at the
frequency $\nu$ of the mean-field oscillation. The other one
describes the frequency of the clusters. This becomes clear when
considering the Fourier amplitudes, corresponding to these peaks, in
the complex plane: At $\omega = \nu$
(Fig.~\ref{fig:Fourier_analysis_cluster}b and e) all local oscillators form a
bunch, while at $\omega \approx \nu/2$
(Fig.~\ref{fig:Fourier_analysis_cluster}c and f) the oscillators arrange into
two clusters, located at the endpoints of the bar visible. Due to the
diffusive coupling, the clusters are connected by an interfacial
region, leading to the intermediate oscillators of the bar.
The fact that the connection of the two clusters crosses the
zero point implies that the borders between them are Ising-type walls.
As in this picture the phase shift between the two clusters is given by $\pi$, we
conclude that at this frequency the clustering takes place. In the
experiments in most cases the cluster frequency is given by approximately
$\nu/2$. This leads to the conclusion that the clusters arise via a
period-doubling bifurcation. Contrarily, in the theory the cluster
frequency can be tuned continuously. For better comparison, we chose
the parameter values such that the frequency also amounts to $\nu/2$.

\clearpage

\subsection{Subclustering}

\begin{figure}[ht]
  \centering
  \includegraphics[width=11cm]{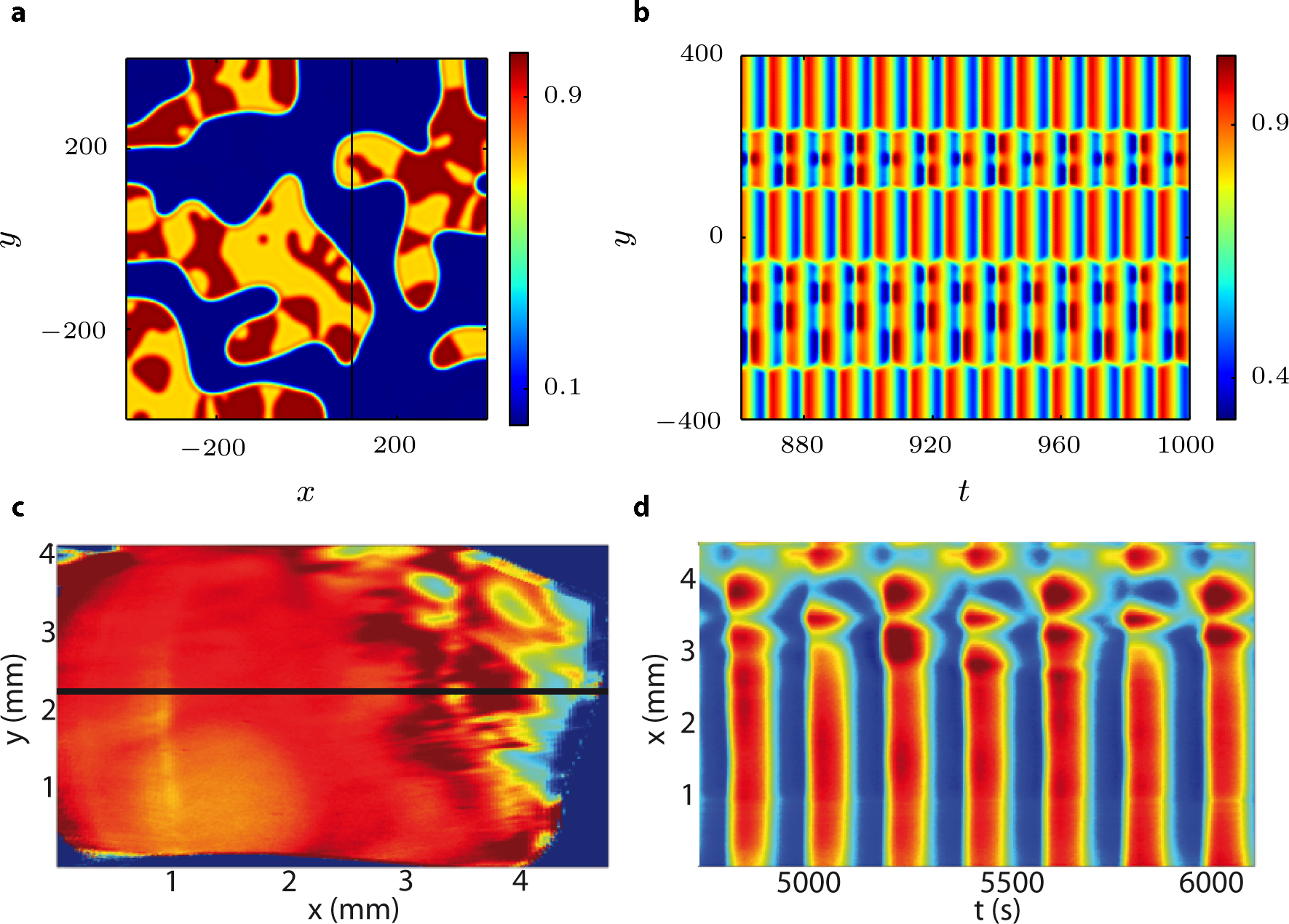}
  \caption{Subclustering in theory (a,b) and experiment (c,d) . Again
    snapshots (a,c) and one-dimensional cuts (b,d) are shown. The
    system splits into two regions, one region being homogeneous and
    one exhibiting two-phase clusters as a substructure. Note that in
    (b) for better visibility $\left| W(y,t) \right|$ is shown. Parameters read: $c_1 = 0.2$, $c_2 = -0.67$, $\nu = 0.1$, $\eta = 0.66$ (simulation) and $c_{\mathrm{F}}=35$ mM, pH=1, $R_{\mathrm{ext}}\cdot A=7.6$ k$\Omega$cm$^2$, $I_{\mathrm{ill}}=0.5$ mW/cm$^2$ (experiment).}
\label{fig:subcluster}
\end{figure}

A symmetry-breaking type of clustering also occurs in our simulations
and experiments. The system again separates into two regions as in the
case of the two-phase clusters, but now one region is homogeneous,
while the other one exhibits two-phase clusters as a
substructure \cite{Schmidt_Chaos_2014}. Such states were also observed in
Refs.~\cite{Kaneko_PhysicaD_1990, Tinsley_Nature_2012}. The results
are depicted in Fig.~\ref{fig:subcluster}. 

In the simulations the two-phase subclusters oscillate at half the
frequency of the main clusters. This indicates that the subclustering
is connected to a period doubling bifurcation.

\subsection{Chimera states}
In the preceding section we demonstrated that the symmetry of the
two-phase cluster pattern can be broken, resulting in a substructure
in one of the two domains. This symmetry-breaking can be even more
dramatic: one of the two domains does not exhibit a coherent
substructure, it rather displays turbulent dynamics. Thus the whole
system separates into two regions, one being synchronized, while the
other one displays incoherent and chaotic oscillations. This
coexistence of synchrony and asynchrony was termed a chimera state
\cite{AbramsStrogatz_PRL_2004} and many theoretical investigations
deal with this topic \cite{Kuramoto_NPCS_2002, AbramsStrogatz_PRL_2004, Shima_PRE_2004, Martens_PRL_2010,
    Sethia_PRL_2008, Abrams_PRL_2008, Schoell_PRL_2011,
    Omelchenko_PRE_2012, Nkomo_PRL_2013, Omelchenko_PRL_2013, Schmidt_Chaos_2014}. Chimera states might be of importance for some peculiar
observations in different disciplines, such as the unihemispheric
sleep of animals \cite{Rattenborg_NBRev_2000, Mathews_Ethology_2006}, the need for synchronized bumps in
otherwise chaotic neuronal networks for signal propagation \cite{Vogels_ARN_2005} and
the existence of turbulent-laminar patterns in a Couette flow
\cite{Barkley_PRL_2005}. They could also be realized experimentally in
chemical, optical, mechanical and electrochemical systems
\cite{Tinsley_Nature_2012, Hagerstrom_Nature_2012, Martens_PNAS_2013,
  Wickramasinghe_PONE_2013, Schmidt_Chaos_2014}.

\begin{figure}[ht]
  \centering
  \includegraphics[width=11cm]{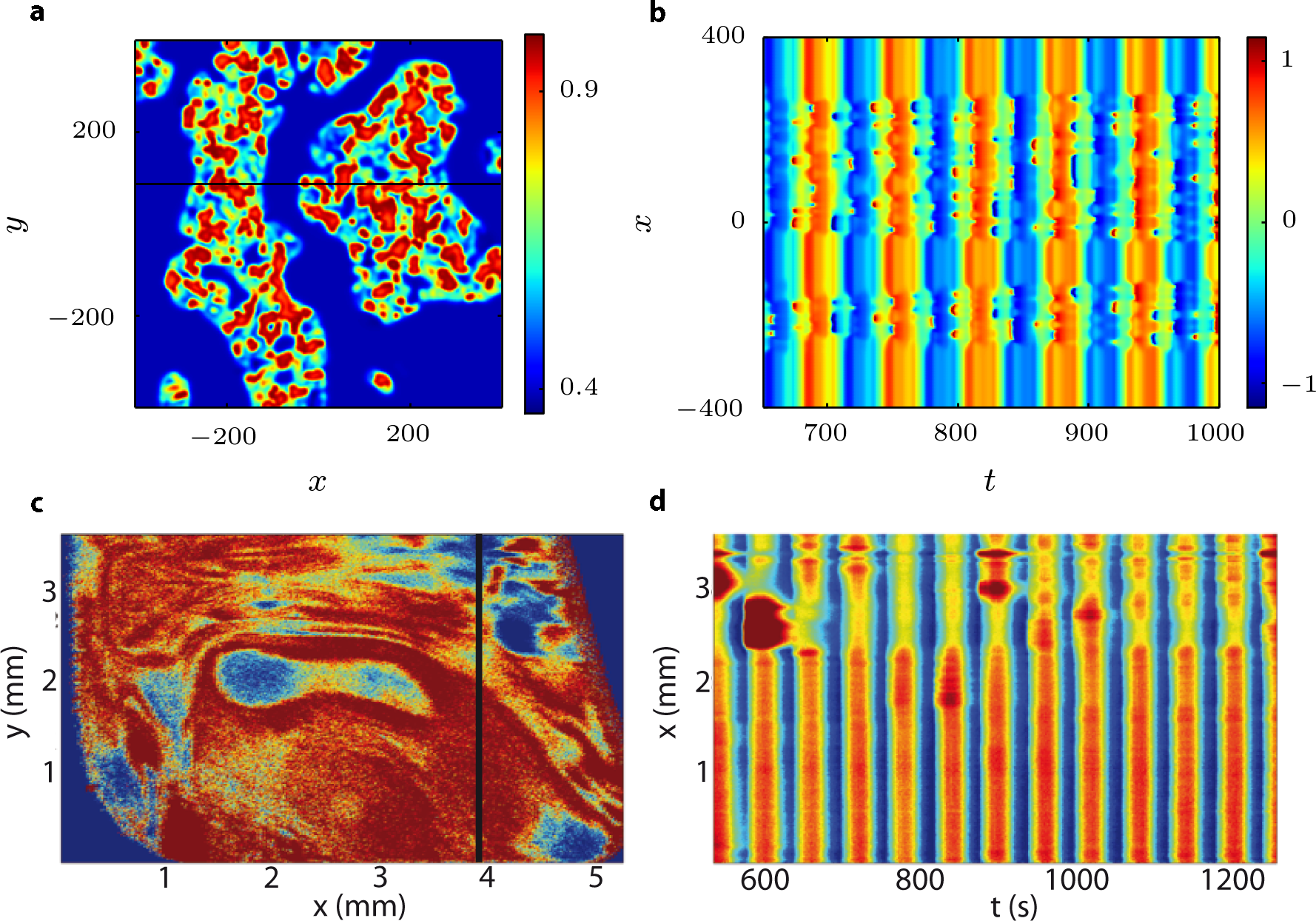}
  \caption{Chimera states in theory (a,b) and experiment
    (c,d). Snapshots (a,c) and one-dimensional cuts (b,d) are
    shown. In both, the simulation and the experimental pattern, the
    synchronized and turbulent regions can be clearly distinguished. Parameters read: $c_1 = 0.2$, $c_2 = -0.58$, $\nu = 0.1$, $\eta = 0.66$ (simulation) and $c_{\mathrm{F}}=50$ mM, pH=3, $R_{\mathrm{ext}}\cdot A=4.5$ k$\Omega$cm$^2$, $I_{\mathrm{ill}}=0.5$ mW/cm$^2$ (experiment).}
\label{fig:chimera}
\end{figure}

In Fig.~\ref{fig:chimera} we present the chimera states found in the
simulations (a,b) and in the experiments (c,d) \cite{Schmidt_Chaos_2014}. 

In both the experiments and the simulations nothing is imposed to induce this
symmetry-breaking. The experimental conditions are kept uniform over
the entire electrode. Furthermore, these patterns form spontaneously,
i.e. no specially prepared initial conditions are required to obtain
them. We could show in Ref.~\cite{Schmidt_Chaos_2014} that these
peculiar dynamics can be traced back to the nonlinear global
coupling. In order to proof this, we performed simulations of Eq.~\eqref{eq:MCGLE} without the
diffusive coupling, which means that we dealt with an ensemble of
Stuart-Landau oscillators coupled solely via the nonlinear global
coupling present in Eq.~\eqref{eq:MCGLE}. Also in this system the
chimera state arises and has the same features as the one presented
here. Thus, contrarily to the convincement in literature that a
nonlocal coupling is indispensable for the emergence of chimera
states, the chimera state found in our simulations
forms under a solely global coupling. The diffusive coupling leads to
the spatial arrangement into synchronized and incoherent regions. In
the experiments a nonlocal contribution of the nonlinear coupling
could not be ruled out yet. However, the striking similarity between
simulations and experiments strongly corroborates the notion that the
nonlinear coupling acts essentially globally.

\subsection{Turbulence}

\begin{figure}[ht]
  \centering
  \includegraphics[width=11cm]{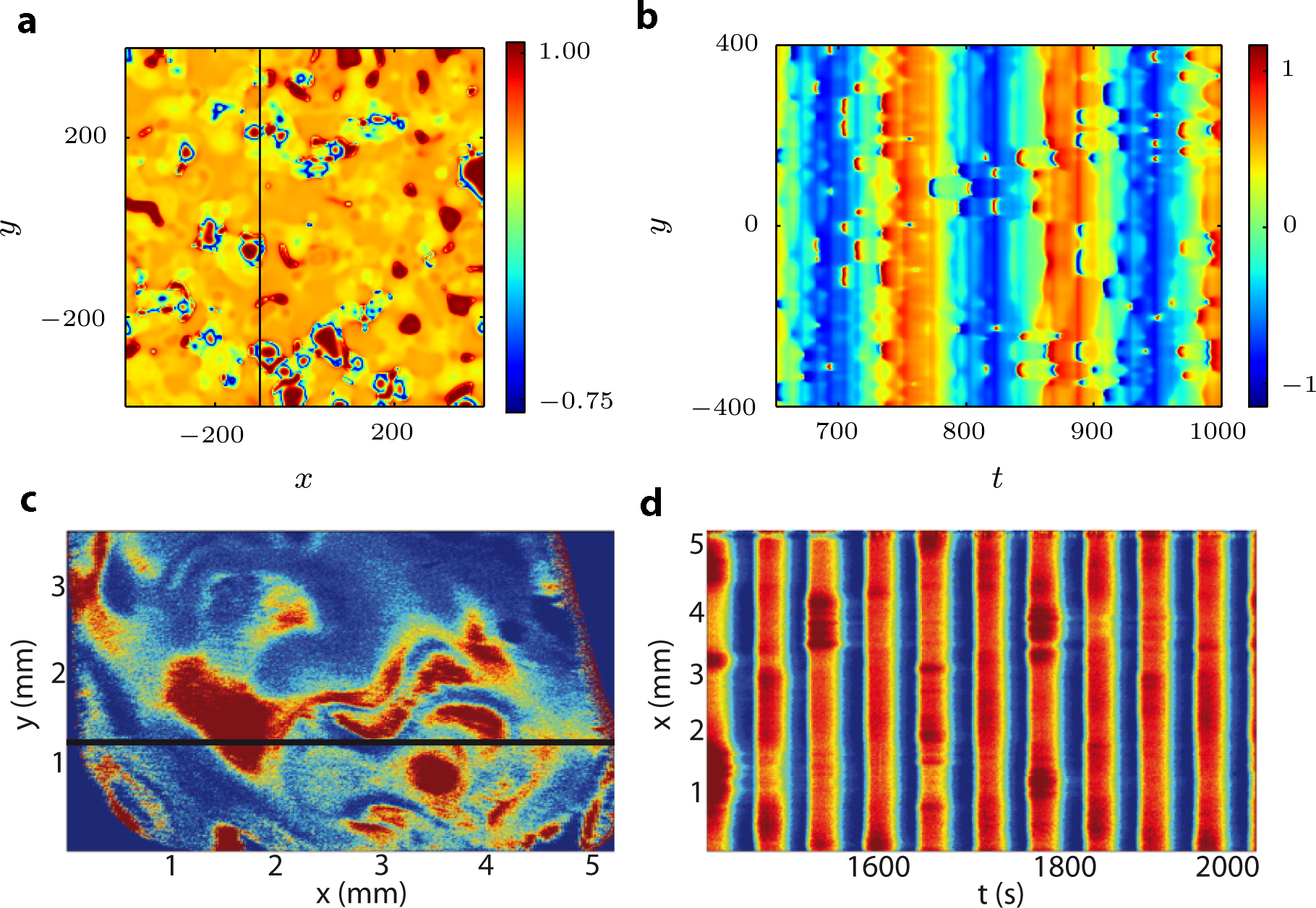}
  \caption{Turbulent dynamics in theory (a,b) and experiment
    (c,d). Snapshots (a,c) and one-dimensional cuts (b,d) are
    shown. The whole system exhibits turbulent dynamics. Parameters read: $c_1 = 0.2$, $c_2 = -0.58$, $\nu = 0.05$, $\eta = 0.66$ (simulation) and $c_{\mathrm{F}}=50$ mM, pH=3, $R_{\mathrm{ext}}\cdot A=6.7$ k$\Omega$cm$^2$, $I_{\mathrm{ill}}=1.0$ mW/cm$^2$ (experiment).}
\label{fig:turbulent}
\end{figure}

As the coexistence of synchrony and turbulence in the chimera state
suggests, we find these states in parameter space between the fully
synchronized and the turbulent states. Therefore, the chimera state is
kind of an mediator between synchrony and turbulence. An experimental example
of the synchronized state is shown in Fig.~\ref{fig:NurHom}, whereas
the turbulent dynamics are presented in Fig.~\ref{fig:turbulent}.

A uniform oscillation is still present in dynamics, but it is
modulated by incoherent and aperiodic oscillations. Similar turbulent
patterns have been found in Ref.~\cite{Vanag_Nature_2000}, where also
localized clusters are discussed, which are reminiscent of the
subclusters presented in Fig.~\ref{fig:subcluster}.

\section{Conclusions \& Outlook}
The oscillatory photoelectrodissolution of n-type silicon is a
convenient experimental model system
exhibiting a wide variety of dynamical states. Most importantly, the
formation of many qualitatively different patterns is observed while
the spatial average of the oscillations is simple periodic. This
behavior can be well reproduced in theoretical simulations using a
general normal form approach close to the Hopf bifurcation adjusted to
capture the mean-field oscillation. 

In the experiments, the points at the
elecrode surface are coupled by a linear
global coupling and a nonlinear coupling both linked to the
total current through the electrode surface. In the theoretical
modelling this is reflected by the introduction of a nonlinear, purely
global coupling. Although the details of the couplings introduced in
experiment and theory are different, the results show
striking similarities which led us to the conclusion that the
observed phenomena are dynamical in origin and robust. Furthermore, it
strongly suggests that the nonlinear coupling introduced in the
experiments by the insufficient illumination is essentially global and
that, e.g., the diffusion length of the minority charge carriers in
the silicon plays an only minor role. An important point that might
be counterintuitive at first glance, is the nature of the patterns
emerging spontaneously in this purely globally coupled system. They
often consist of distinct regions showing remarkably different
dynamical behavior, the most astonishing example being the chimera
state. Global couplings occur naturally in oscillatory systems in many
fields of research \cite{Purwins:2010, Krischer_2003,
  Strogatz_PhysicaD_2000, Wacker_JAP_1995}. The variety of dynamical patterns
observed and their spontaneous emergence in the presented experimental
system thus offers intriguing insight into the dynamical possibilities
of a wide variety of systems.

\section*{Acknowledgments}
The authors gratefully acknowledge financial support from the
\textit{Deutsche Forschungsgemeinschaft} (Grant
no. KR1189/12-1), the \textit{Institute for Advanced Study, Technische
  Universit\"{a}t M\"{u}nchen} funded by the German Excellence
Initiative and the cluster of excellence \textit{Nanosystems
  Initiative Munich (NIM)}.


\end{document}